\documentclass[a4paper]{spie}

\usepackage{amsmath,amsfonts,amssymb}
\usepackage{bm}   
\usepackage{graphicx}
\usepackage[colorlinks=true, allcolors=blue]{hyperref}
\usepackage{astro_bib_macro}
\usepackage{ulem}

\def\Var{{\textrm{Var}}\,}

\DeclareMathOperator{\tr}{tr}

\usepackage{booktabs}

\usepackage{listings}

\title{Predicting contrast sensitivity to segmented aperture misalignment modes for the HiCAT testbed}

\author{Iva Laginja\supit{a}\supit{b}\supit{c}, R\'{e}mi Soummer\supit{c}, Laurent M. Mugnier\supit{a}, Laurent Pueyo\supit{c}, Jean-Fran\c{c}ois Sauvage\supit{a}\supit{b}, Lucie Leboulleux\supit{d}, Laura Coyle\supit{e}, J. Scott Knight\supit{e}, Marshall D. Perrin\supit{c}, Scott D. Will\supit{c}\supit{f}, James Noss\supit{c}, Keira J. Brooks\supit{c}, Julia Fowler\supit{c}\supit{g}
\skiplinehalf
\supit{a} DOTA, ONERA, Universit\'e Paris Saclay, F-92322 Ch\^{a}tillon, France\\
\supit{b} Aix Marseille Universit\'{e}, CNRS, LAM (Laboratoire d'Astrophysique de Marseille) UMR 7326, 13388 Marseille, France\\
\supit{c} Space Telescope Science Institute, Baltimore, MD 21218, USA\\
\supit{d} LESIA, Observatoire de Paris, Universit\'{e} PSL, CNRS, 92195 Meudon, France\\
\supit{e} Ball Aerospace \& Technologies Corp., Boulder, CO 80301, USA\\
\supit{f} The Institute of Optics, University of Rochester, Rochester, NY 14627 USA\\
\supit{g} University of California Santa Cruz, 1156 High Street, Santa Cruz, CA 95064, USA
}

\authorinfo{Further author information, send correspondence to Iva Laginja: E-mail: iva.laginja@lam.fr}

\begin{document} 
\maketitle

\begin{abstract}

This paper presents the setup for empirical validations of the Pair-based Analytical model for Segmented Telescope Imaging from Space (PASTIS) tolerancing model for segmented coronagraphy. We show the hardware configuration of the High-contrast imager for Complex Aperture Telescopes (HiCAT) testbed on which these experiments will be conducted at an intermediate contrast regime between $10^{-6}$ and $10^{-8}$. We describe the optical performance of the testbed with a classical Lyot coronagraph and describe the recent hardware upgrade to a segmented mode, using an IrisAO segmented deformable mirror. Implementing experiments on HiCAT is made easy through its top-level control infrastructure that uses the same code base to run on the real testbed, or to invoke the optical simulator. The experiments presented in this paper are run on the HiCAT testbed emulator, which makes them ready to be performed on actual hardware. We show results of three experiments with results from the emulator, with the goal to demonstrate PASTIS on hardware next. We measure the testbed PASTIS matrix, and validate the PASTIS analytical propagation model by comparing its contrast predictions to simulator results. We perform the tolerancing analysis on the optical eigenmodes (PASTIS modes) and on independent segments, then validate these results in respective experiments. This work prepares and enables the experimental validation of the analytical segment-based tolerancing model for segmented aperture coronagraphy with the specific application to the HiCAT testbed.

\end{abstract}


\section{INTRODUCTION}
\label{sec:introduction}

Imaging an Earth-like exoplanet is one of the most highly sought science goals in astronomy today\cite{Roberge2019, Guyon2012}. While there are several viable techniques for giant gaseous exoplanet detection, the method of direct imaging allows us to access the light from these extra-solar systems directly, enabling a generalized spectral study of the planetary atmosphere’s chemical composition and search for the presence of biomarkers. In the case of Earth-like planets the goal is more ambitious as the observatories will need to reach contrast levels (planet to star flux ratios) of at least $10^{-10}$, at a separation of only $\sim 0.1$ arcsec from the star, compared to $10^{-6}$ at $\sim 0.3$ arcsec for faint gas giants. Telescopes providing these capabilities will require large collecting areas, and they will most likely be realized with segmented primary mirrors, both in space and on the ground. Currently, the favored method to achieve these extreme high contrast levels are dedicated instruments, which strongly attenuate the on-axis star light while preserving the off-axis planet light\cite{Martinez2008}. These systems are very sensitive to residual wavefront aberrations, which generate speckles of light in the imaging focal plane that can be mistaken for planets. This is why coronagraphy needs to be combined with wavefront sensing and active control (WFS\&C) to create a zone of deep contrast in the final image, a dark hole (DH). These long-term goals will be achieved from space by missions such as the Habitable Exoplanet Observatory\cite{Gaudi2020} (HabEx) and Large UV Optical InfraRed Surveyor\cite{TheLUVOIRTeam2019, Bolcar2019} (LUVOIR) currently under consideration by the NASA Decadal Survey, with the Nancy Grace Roman Space Telescope (RST)\cite{Krist2015} working towards shorter-term demonstrations at more moderate contrast levels ($\sim 10^{-9}$).

Due to the high sensitivity of coronagraphs to wavefront errors (WFE), a careful analysis of all aberration sources and their impact on contrast is paramount when designing an imaging system. In particular, cophasing errors of the primary mirror segments like in the case of LUVOIR will significantly contribute to these aberrations, which have a degrading effect on the coronagraph contrast. Previous studies have determined cophasing requirements for large segmented apertures to be on the order of 10~pm over the full pupil\cite{Juanola-Parramon2019}, with the objective of maintaining a DH contrast of $10^{-10}$. Instead of defining WFE tolerances over the entire telescope pupil, the Pair-based Analytical model for Segmented Telescope Imaging from Space (PASTIS)\cite{Leboulleux2017, Leboulleux2018spie, Leboulleux2018jatis, Laginja2019} allows us to derive segment-specific requirements, either for each segment individually when they are statistically independent from each other, or in the form of a segment covariance matrix when there are correlations between them\cite{Laginja2020}.

The next step is to perform an experimental validation of these segment-level tolerances on a segmented mirror of a high contrast instrument in a laboratory setting to seek confirmation that they indeed yield the coronagraphic contrast they were calculated for. In this paper, we present the adaptation of the theoretical results to an application on the High-contrast imager for Complex Aperture Telescopes (HiCAT) testbed\cite{N'Diaye2015,Soummer2018}. In order to derive realistic requirements that can be validated with a given instrument, we set up a specific hardware configuration that is capable of operating at a sufficiently deep dark hole contrast and include an IrisAO segmented deformable mirror (DM) in the optical setup. We define a target contrast with a corresponding WFE perturbation amplitude that is compatible with the performance of the installed segmented DM and carry out emulated experiments on the integrated HiCAT simulator, which are ready to run on the hardware as-is. 

In Sec.~\ref{sec:recall-pastis} we recall the most important points about the PASTIS forward model and its use in WFE tolerancing. In Sec.~\ref{sec:hicat-testbed-simulator} we describe the HiCAT project, its synthetic testbed emulator including an optical end-to-end simulator, and the testbed configuration used for the presented experiments. In Sec.~\ref{sec:results-hicat} we show the tolerancing results and their validations performed on the HiCAT emulator and in Sec.~\ref{sec:CONCLUSIONS} we report our conclusions and considerations for the upcoming hardware validations.

Note that the objective of the PASTIS model is the spatially averaged contrast in the dark hole (normalized to the peak of the direct image), which is what we refer to as ``contrast" throughout this paper. We also stress that we differentiate between this spatially averaged dark hole intensity, the ``average DH contrast", and a statistical mean (expectation value) of this averaged contrast over many optical propagations, the statistical ``mean contrast".


\section{PASTIS model summary}
\label{sec:recall-pastis}

 PASTIS\cite{Leboulleux2017, Leboulleux2018spie, Leboulleux2018jatis, Laginja2019, Laginja2020} is an analytical forward model that directly calculates the average contrast over the DH, a scalar quantity, from a set of segment-level aberration amplitudes in the pupil plane. Central to this model is the PASTIS matrix $M$, which describes the contrast contributions of an aberrated segment pair. This matrix is a full representation of a given high contrast optical system, including its coronagraph and primary mirror geometry. Combined with the separately defined mechanical properties of the segmented aperture, expressed as a segment covariance matrix $C_a$, it can be used to calculate the expected mean contrast of a particular instrument, no matter the nature of the correlation between the segments. Additionally, it allows us to calculate the contrast variance of this distribution. The ability to calculate this information (independent segment requirements, analytical mean contrast and variance derivation) describes the statistical response of a segmented coronagraph to segment-level cophasing errors. In this paper, we use the semi-analytical development of the PASTIS model\cite{Laginja2019, Laginja2020} to calculate a PASTIS matrix with an emulated HiCAT testbed, which we use for further analysis. In the following section, we present a brief summary of the tolerancing model.

Representing optical aberrations on a segmented telescope with local Zernike modes, we can expand the phase aberrations on the segmented pupil $\phi_s$ as a sum of such segment-level polynomials \cite[Eq.~1]{Laginja2020}. We then represent the total phase in the pupil with two terms, one best-contrast phase solution, $\phi_{DH}$, and the segmented perturbation, $\phi_s$. Under the assumption of the small aberration regime for $\phi_s$, we can express the average contrast in the DH as a matrix multiplication:
    \begin{equation}
    c = c_0 + \mathbf{a}^T M \mathbf{a},
    \label{eq:pastis-equation}
    \end{equation}
where $c$ is the spatial average contrast in the dark hole, $c_0$ the coronagraph floor (i.e. the average contrast in the dark hole at best contrast with $\phi_{DH}$, in the absence of additional phase perturbations), $M$ is the PASTIS matrix with elements $m_{ij}$, $\mathbf{a}$ is the aberration vector of the local Zernike coefficients on all discrete $n_{seg}$ segments and $\mathbf{a}^T$ its transpose. It was previously shown that the average DH contrast with varying DH solutions can always be expressed as a quadratic function of a segmented phase perturbation under an appropriate change of variable\cite[Eq.~4]{Laginja2020}.

The PASTIS matrix represents the contrast contribution to the DH average contrast by each aberrated segment pair in the pupil, $c_{ij}$. In its semi-analytic calculation, we first calculate the DH average contrast with an end-to-end simulator, before computing the PASTIS matrix elements $m_{ij}$ analytically with\cite[Eqs.~16 and ~17]{Laginja2020}, which for the matrix diagonal is:
    \begin{equation}
     m_{ii} = \frac{c_{ii} - c_0}{a_c^2},
     \label{eq:diagonal-elements}
    \end{equation}
and for the off-diagonal matrix elements:
    \begin{equation}
     m_{ij} = \frac{c_{ij} + c_0 - c_{ii} - c_{jj}}{2 a_c^2},
     \label{eq:off-diagonal-elements}
    \end{equation}
where $a_c$ is the calibration aberration amplitude that is put on each segment in the calculation of the contrast matrix ($c_{ij}$). While the PASTIS matrix elements could equally be calculated from the image plane electric field directly, this is not a viable option in empirical tests, since the electric field can only be estimated, while the DH intensity can be directly measured.

Once the PASTIS matrix is established, we can calculate its eigenmodes $\mathbf{u}_p$ and eigenvalues $\lambda_p$ by means of an eigendecomposition. These modes form an orthonormal basis with a deterministic effect on the average DH contrast, where each mode has a contrast contribution $c_p$. This contrast contribution can be defined for each mode individually, which allows us to derive the required mode weight $b_p$ as a WFE rms\cite[Eq.~25]{Laginja2020}, to attain this contrast. To illustrate one particular way of allocating contrast contributions to each mode, we can calculate the mode weights corresponding specifically to a uniform contrast contribution of the overall target contrast $c_t$ over all modes, $c_p = (c_t - c_0)/n_{modes}$\cite[Eq.~26]{Laginja2020}:
    \begin{equation}
    \widetilde{b_p} = \sqrt{\frac{c_t - c_0}{n_{modes} \cdot \lambda_p}},
    \label{eq:calc-sigma-uniform}
    \end{equation}
where $\widetilde{b_p}$ is the particular weight of a mode with index $p$ in the case of a uniform contrast allocation across all modes.

It follows that the mode weights $b_p$ have the statistical meaning of a standard deviation in a zero-mean normal distribution\cite[Eq.~28]{Laginja2020}, describing the resulting average DH contrast over many realizations of segmented pupil aberrations. In the same way, we can derive segment-level WFE requirements when assuming independent segments, where the tolerances are given in the form of a standard deviation $\mu_k$ per segment $k$, assuming that all segments contribute equally to the final contrast\cite[Eq.~37]{Laginja2020}:
    \begin{equation}
    \mu_k^2  = \frac{c_t -  c_0}{n_{seg} m_{kk}},
    \label{eq:single-mus}
    \end{equation}
where $c_t = \langle c \rangle$ is the targeted mean contrast over many realizations, $c_0$ the coronagraph floor, $n_{seg}$ the total number of segments and $m_{kk}$ the diagonal elements of the PASTIS matrix.

While Eq.~\ref{eq:single-mus} assumes independent segments, which can be expressed with a diagonal segment covariance matrix, $C_a$, the PASTIS matrix allows us to calculate the expected mean of the average DH contrast and its variance directly from any given (correlated or uncorrelated) segment covariance matrix, even if non-diagonal, with\cite[Eqs.~31 and 33]{Laginja2020}:
    \begin{equation}
    \langle c \rangle = c_0 + \tr(M C_a),
    \label{eq:avg-contrast-from-trace}
    \end{equation}
where $\tr$ denotes a trace, and:
    \begin{equation}
    \Var (c) = 2 \tr [(M C_a)^2].
    \label{eq:var-of-c}
    \end{equation}
The two equations above allow us to calculate these two integral quantities directly from the optical properties of the instrument, described by $M$, and the mechanical correlations of the segments, captured by $C_a$, without having to run any Monte Carlo simulations, which allows for an analytical prediction of the mean DH contrast.


\section{The HiCAT testbed and experiment emulation}
\label{sec:hicat-testbed-simulator}

\subsection{The HiCAT project}
\label{subsec:hicat-general-project}

The HiCAT testbed  (High contrast imager for Complex Aperture Telescopes\cite{N'Diaye2013,N'Diaye2014,N'Diaye2015,Leboulleux2017HiCAT,Leboulleux2016,Moriarty2018,Soummer2018}) is  dedicated to a LUVOIR-type coronagraphic demonstration with on-axis segmented apertures. The project is targeting system-level experiments in ambient conditions that can happen before demonstrations in vacuum, for example at the Decadal Survery Testbed (DST)\cite{Patterson2019} located at the Jet Propulsion Laboratory. 

The HiCAT testbed\cite[Fig.~1]{Soummer2018} incorporates three DMs: two Boston Micromachines 952-actuator micro electro-mechanical (MEMS) ``kilo-DMs" and an IrisAO PTT111L 37-element hexagonally-segmented DM\cite{Helmbrecht2013}. A central obstruction and support structures, as well as any arbitrarily shaped pupil masks can be added in the first pupil plane, using a laser-cut or etched transmissive mask. The segmented DM has a calibrated surface error of $9~nm$ rms\cite[Fig.~3]{Soummer2018}, with very high open-loop repeatability, which makes it directly suitable for the high contrast goals of HiCAT. The HiCAT telescope simulator therefore is truly segmented with the ability to add real co-phasing wavefront errors and introduce temporal drifts for dynamical studies. The workhorse coronagraph configuration for HiCAT is an Apodized Pupil Lyot Coronagraph (APLC) \cite{N'Diaye2015, N'Diaye2016, Zimmerman2016} that includes apodizers manufactured using carbon nanotubes grown on a customized catalyst for the black areas, and protected silver or gold for the reflective areas. These masks are manufactured by Advanced Nanophotonics Inc.\cite{Hagopian2010} These optical components are mounted on easily interchangeable bonding cells that allow a fast change between different designs and optics in the apodizer pupil plane, or to swap in a high-quality flat mirror to put the testbed in a classical Lyot coronagraph (CLC) configuration. In previous experiments, the IrisAO was installed in CLC mode for experimental validations of coronagraphic focal plane wavefront sensing on a segmented aperture\cite{Leboulleux2020}. The testbed also includes a fast tip-tilt system, and an extensive supporting metrology suite (custom interferometric metrology for critical hardware elements, phase retrieval channel, theodolites, cameras) that allows for fast and precise component changes. The phase retrieval camera can be used to measure the wavefront at a focal plane mask (FPM) proxy location by introducing a high-quality flat mirror into the beam \cite{Brady2018,Brady2019}. A Zernike wavefront sensor has been assembled as part of the low-order wavefront sensor, using the light rejected by the FPM.\cite{Pourcelot2020}

\subsection{HiCAT testbed emulator and controls}
\label{subsec:hicat-emulator}

The high-level testbed control system of HiCAT is unique due to its dual-mode operation setup. Coded purely in Python, the same code base is used to control either the actual testbed hardware or to use the moderately high fidelity optical simulator for HiCAT. The latter includes a full simulation of the hardware control interfaces, which makes it a complete ``synthetic testbed". In practice, this means that the \textit{exact} scripts that are used to run testbed simulations can also be run on the actual testbed, allowing us to first test experiments on the simulator, and then work right out of the box on the hardware. This works with identical commands to the motor controllers, DMs, and all hardware components, using the same pipeline for data processing and producing the same output data products. It is this ``emulated testbed" that we use to produce the results presented in this paper.

The overall control system and software architecture is object-oriented and modular, and is hosted on GitHub, while also deploying automatic testing and continuous integration. The hardware controls have been abstracted in the public CATKit\footnote{https://github.com/spacetelescope/catkit} Python package, which provides the interface to all our hardware components, e.g. from Boston Micromachines, IrisAO, Thorlabs, Newport and others. The HiCAT optical simulator uses the POPPY Python Fourier optics toolkit\cite{Perrin2016poppy, Perrin2012poppy}, mixing Fraunhofer and Fresnel models, including a fast semi-analytic  coronagraph propagation at high resolution. The simulator was originally constructed from the theoretical testbed optical design and later refined with empirical calibrations and measured optical alignments. It enables model-based control algorithms with a Jacobian calculated on the simulator, e.g. for pair-wise probing\cite{Groff2016} and stroke minimization\cite{Pueyo2009, Mazoyer2018a, Mazoyer2018b}.

\subsection{HiCAT testbed configuration for experimental validation of PASTIS}
\label{subsec:hicat-for-pastis}

While the HiCAT APLC is designed to provide a superior performance on a segmented aperture compared to the simpler CLC, the pupil plane apodization of this coronagraph causes a lot of the aperture segments to be highly concealed (see \cite[Fig.~7]{Soummer2018}). This will have a direct impact on the segment-level tolerancing as described by the PASTIS model\cite[Sec.~6]{Laginja2020}. Moreover, the pupil apodization and the FPM filtering display competing effects in the tolerancing, which is why we chose to perform the empirical validation of PASTIS on a CLC configuration of HiCAT. This is the setup in which the HiCAT hardware has been operating over the past months, except that there was no segmented DM installed. Adding the IrisAO puts HiCAT into a ``segmented CLC mode", which includes a pupil mask that traces the segmented IrisAO outline, and its largest diameter is slightly undersized from that of a circumscribed circle around the IrisAO, with diameter $D_{pup}$. HiCAT hosts one in-pupil and one out-of-pupil 1k Boston continuous DM, an FPM with a radius of 8.56 $\lambda/D_{pup}$ and a circular, unobscured Lyot stop. Its diameter $D_{LS}$ is undersized by two percent with respect to the inscribed circle of the IrisAO (projected in the Lyot plane), keeping its edges within the controllable area of the segmented mirror (and the hexagonally outlined pupil mask, see Fig.~\ref{fig:hicat_pupil_overlaps}, left), which puts its size at 81\% of $D_{pup}$. With the WFS\&C strategy on HiCAT, using pair-wise estimation and stroke minimization, the unsegmented version of this CLC setup (with a circular pupil instead of the outline mask, with diameter $D_{pup}$) achieved a contrast of $4\times10^{-8}$ on hardware, in monochromatic light at $640~nm$, and $2\times10^{-7}$ in 10\% broadband light, in a $360^{\circ}$ DH from 6-11 $\lambda/D_{LS}$ (where the outer working angle is defined by the highest spatial frequency controllable by the continuous DMs).

\begin{figure}
   \begin{center}
   \begin{tabular}{c}
   \includegraphics[width = 12cm]{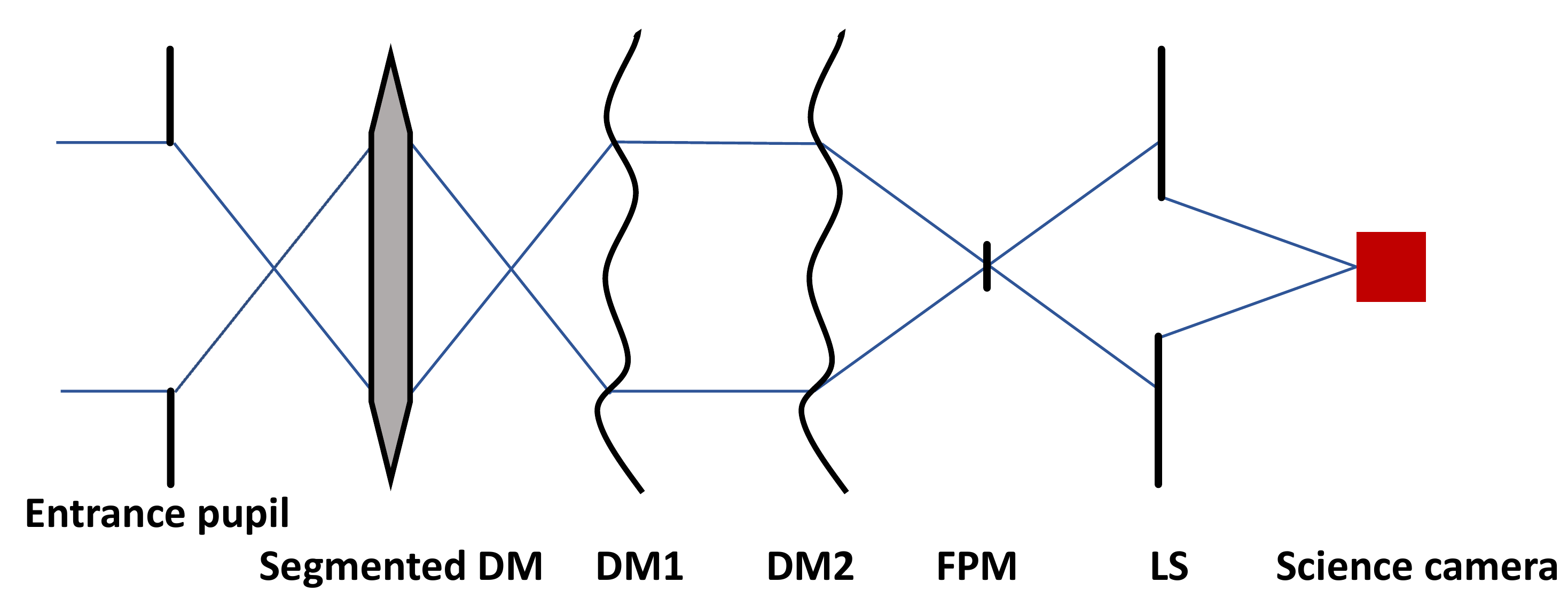}
   \end{tabular}
   \end{center}
   \caption[HiCAT layout for PASTIS] 
   {\label{fig:hicat_layout_pastis} 
    HiCAT optical configuration used for PASTIS experiments, here shown with transmissive optics for simplicity. The entrance pupil is a custom shaped mask tracing the outline of the segmented DM in a consecutive pupil plane. Of the two continuous deformable mirrors, DM1 is in a pupil plane, and DM2 is located out of pupil. The focal plane mask (FPM) and Lyot stop (LS) build the classical Lyot coronagraph setup.}
   \end{figure}

\begin{figure}
   \begin{center}
   \begin{tabular}{c}
   \includegraphics[width = 12cm]{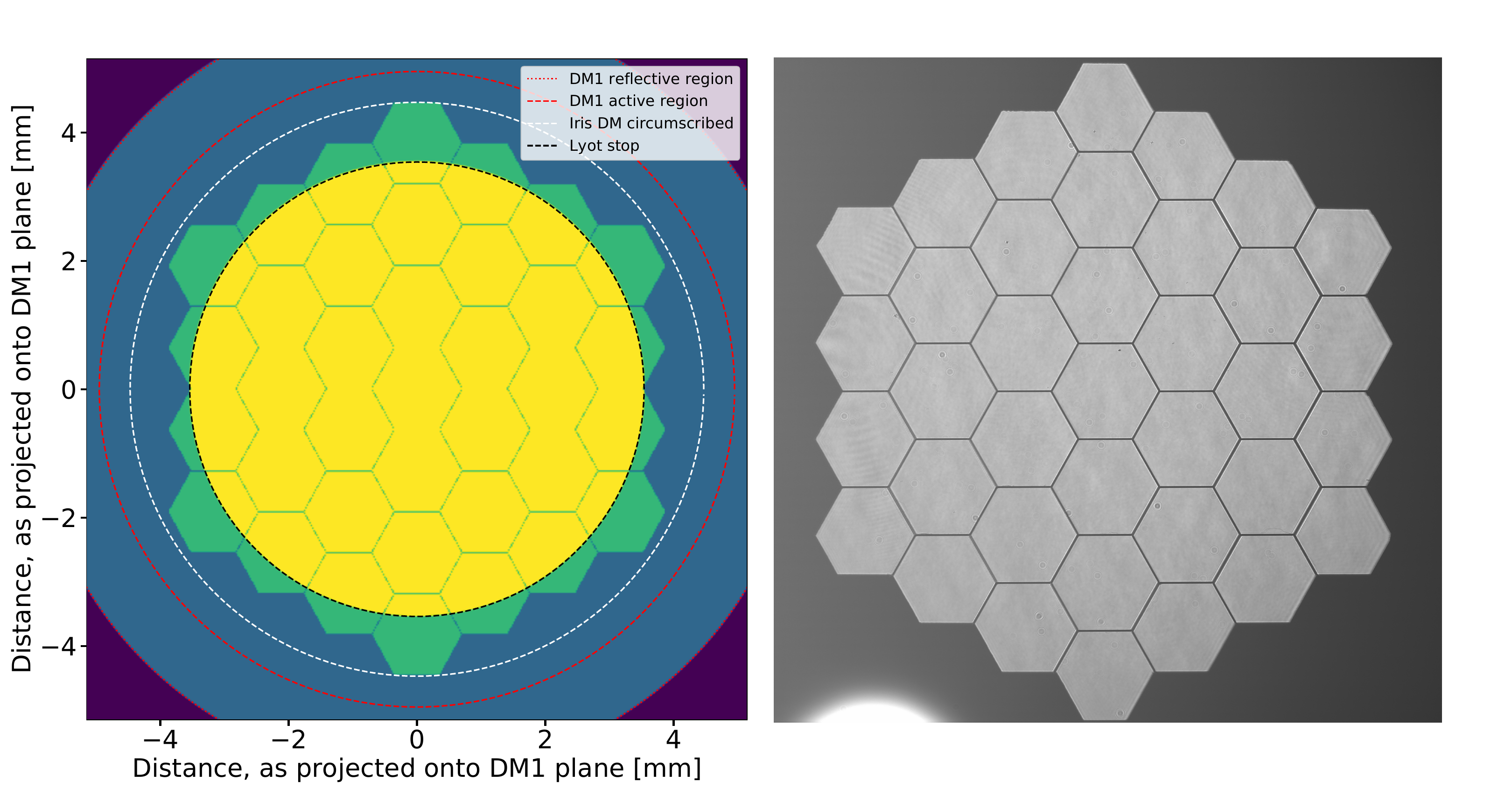}
   \end{tabular}
   \end{center}
   \caption[HiCAT pupil overlaps and hardware pupil image] 
   {\label{fig:hicat_pupil_overlaps} 
    \textit{Left:} Overlapping pupils in the HiCAT segmented CLC configuration used for the PASTIS experiments. The entrance pupil mask traces the outline of the IrisAO, preventing the illumination of areas outside of the controllable segments. The Lyot stop is sized such that its edges stay within the controllable outline of the IrisAO (yellow circle). The entrance pupil diameter $D_{pup}$ is defined as the circumscribed circle around the IrisAO (dashed white). \textit{Right:} Measured pupil image in a pupil plane before the two continuous DMs and the Lyot stop, showing the IrisAO segments and the pupil mask outlining the segmented DM. Note the slightly undersized outline, which results in somewhat irregular hexagons at the edges. The bright spot of light at the bottom comes from the metrology suite that was mounted at the time of the exposure and it does not impact the HiCAT optical performance.}
   \end{figure}

To perform our experimental validations, we have recently installed a 37-segment IrisAO PTT111L\cite{Helmbrecht2016} DM on HiCAT (see Fig.~\ref{fig:hicat_pupil_overlaps}, right) together with said pupil mask that traces its outline, in order to prevent any illumination of the DM surface beyond the controllable segments, which completes the configuration for our empirical validations. This optical testbed layout can be seen in Fig.~\ref{fig:hicat_layout_pastis}.
The segments of the IrisAO are each controllable in piston, tip and tilt, with a maximum stroke of $5~\mu m$. Using a 14bit controller, we assume close to perfect linear actuators\cite{Helmbrecht2007}, resulting in the smallest control step per actuator of $0.3~nm$, excluding any noise. This control step will drive the contrast level we tolerance to in Sec.~\ref{sec:results-hicat}.

\begin{figure}
   \begin{center}
   \begin{tabular}{c}
   \includegraphics[width = 16cm]{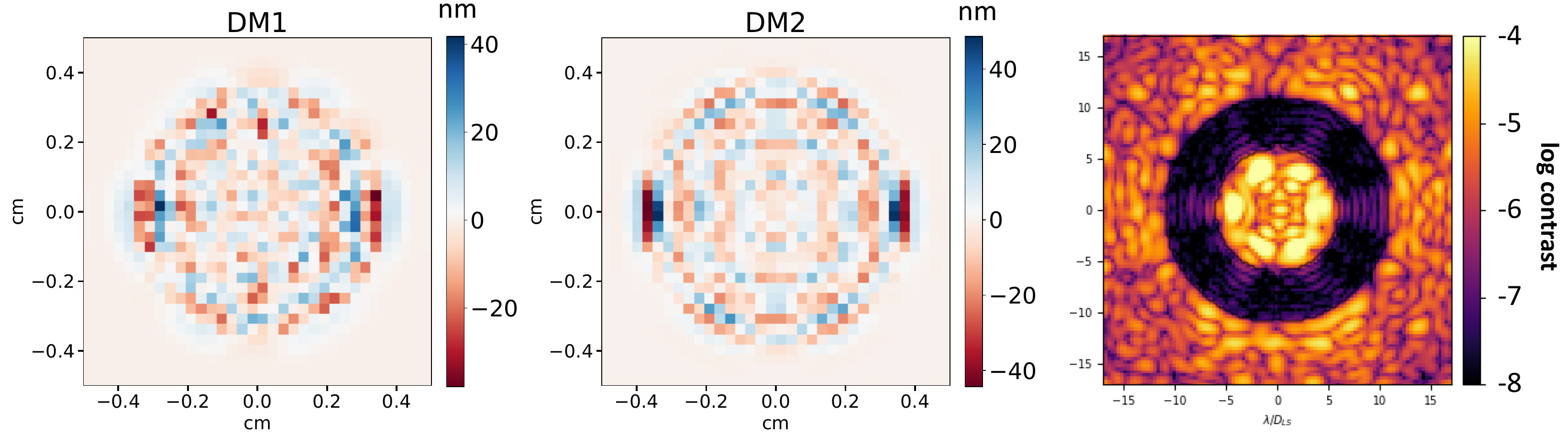}
   \end{tabular}
   \end{center}
   \caption[Baseline DH] 
   {\label{fig:dms-and-dh} 
    DM maps applied to continuous DM1 (in-pupil, left) and DM2 (out-of-pupil, center) to obtain the $360^{\circ}$ DH solution on the HiCAT emulator (right), calculated by 10 iterations of pair-wise sensing and stroke minimization, which yields an average contrast of $5.7 \times 10^{-8}$. The segmented DM in this setup has a static co-phasing WFE of $10~nm$ applied to it, which gets corrected with the WFS\&C loop. We include these DM solutions into the definition of our coronagraph when working with PASTIS.}
   \end{figure}

With this segmented CLC setup on the HiCAT emulator, without any WFS\&C and with a flat segmented DM, the average contrast in the DH area is $1.2 \times 10^{-5}$. In order to place the coronagraph floor of the testbed into a higher contrast regime, we deploy an iterative WFS\&C loop that uses pair-wise sensing to estimate the E-field in each iteration, followed by a control step with both DMs as calculated by the stroke minimization algorithm, replicating the result achieved on hardware in the unsegmented configuration (see above). The IrisAO is kept at its best flat position throughout. We chose to stop this control loop after 10 iterations, which is when we reach an average DH contrast of $5.7 \times 10^{-8}$, comparable to the best contrast HiCAT can reach on the real testbed in an unsegmented CLC configuration, and in monochromatic light. The DM solutions and focal plane image from this simulated experiment are displayed in Fig.~\ref{fig:dms-and-dh}.
The WFS\&C solution shown in Fig.~\ref{fig:dms-and-dh} are included in the static coronagraph contribution of the PASTIS model\cite[Eq.~12]{Laginja2020}, which allows us to work around an improved best contrast solution compared to the coronagraph without deploying a DH algorithm. This sets our coronagraph floor that we use in the emulated PASTIS experiments on HiCAT in Sec.~\ref{sec:results-hicat} to an initial $c_0 = 5.7 \times 10^{-8}$.


\section{Results on the HiCAT emulator}
\label{sec:results-hicat}

The experimental validations of the PASTIS model tolerancing predictions seek to confirm that a set of derived PASTIS mode, and per-segment WFE requirements indeed yield the average DH contrast they were calculated for. For this purpose, we are preparing these experiments on the HiCAT testbed, making use of the ``synthetic testbed" mode of the hardware controls to write the experiments and to estimate sensitivity levels of the individual optical components in the presence of realistic residual WFE on the testbed. These experiments include all preparation steps needed for the anticipated hardware runs and they are expected to run ``out of the box" on the testbed hardware.

In this section, we present the results of these emulated validation experiments. We use the testbed configuration described in Sec.~\ref{subsec:hicat-for-pastis}, using a monochromatic light source at $640~nm$. Before running an experiment, we apply the DM solutions for the DH shown in Fig.~\ref{fig:dms-and-dh} in order to use the DH contrast from that stroke minimization solution as the coronagraph floor, initially $c_0 = 5.7 \times 10^{-8}$. During hardware operations, there will be a temporal contrast drift due to environmental changes, mostly driven by the changing temperature and humidity in the lab, as well as image jitter caused by the air vents injecting dry air into the testbed enclosure. The installation of diffusers on these vents has lead to a great improvement on image stability, and HiCAT is able to recover the same level of contrast in the range of $5-6\times10^{-8}$ by applying an open-loop DH solution up to two weeks after the experiment. The residual jitter will likely result in a systematic, though small, error in the results.

The first experiment is the measurement of a PASTIS matrix, which will be used to calculate the mode and segment requirements for a given target contrast. As described in Sec.~\ref{subsec:hicat-for-pastis}, the least significant bit (LSB) of the IrisAO controller allows for a minimal movement of $0.3~nm$ of a single IrisAO segment, barring any noise. In order to minimize limitations by the LSB, we chose here a conservative target contrast of $c_t = 10^{-6}$, which results in a standard deviation for the segment requirements larger than $4~nm$ (see Sec.~\ref{subsec:segment-tolerances}). Since these tolerances are drawn from a zero-mean distribution, some random WFE realizations in Sec.~\ref{subsec:segment-tolerances} will still be truncated to zero due to the LSB limit, especially when taking into account additional controller noise, but with larger standard deviations we are increasing the fraction of realizations above that limit.

In the following, we present the results of three experiments for the validations of the PASTIS tolerancing model: (1) measuring an empirical PASTIS matrix and validating the PASTIS forward model with a ``hockey stick curve" experiment (see \cite[Fig.~4]{Laginja2020}), (2) measuring the cumulative contrast of the modes obtained from the testbed PASTIS matrix by tolerancing all modes to a uniform contrast contribution, and (3) performing a Monte Carlo experiment for the validation of the calculated segment requirements, where we propagate random WFE maps drawn from the tolerancing prescription for independent segments calculated with PASTIS to measure the resulting DH average contrasts.

\subsection{PASTIS matrix measurement and validation}
\label{subsec:measured-pastis-matrix}

The PASTIS matrix is a pair-wise influence matrix, linking segment aberrations to the average contrast in the coronagraphic DH. Having the scalar quantity of the average DH contrast as its objective, it has the advantage that no prior knowledge of the electric field is required in order to construct it. This means that measuring an empirical PASTIS matrix is much faster compared to measuring an empirical electric field Jacobian, since its calculation does not introduce overheads that usually come with E-field estimation methods (e.g. measuring probe images in the pair-wise estimator). Moreover, there will be no estimation error in the matrix result, and measuring an empirical Jacobian has not been successfully done to date. One aspect to consider is that the measurement time for the PASTIS matrix scales roughly with the square of the number of segments ($n_{seg}$) divided by two, rather than linearly with the number of segments:
    \begin{equation}
    n_{meas} = \frac{n_{seg}^2 + n_{seg}}{2},
    \label{eq:number-of-measurements}
    \end{equation}
where $n_{meas}$ is the total number of measurements required for the construction of the PASTIS matrix. We divide by $2$ because the matrix is symmetrical, and include the matrix diagonal by adding $n_{seg}$. On the 37-segment HiCAT pupil, this requires only $n_{meas} = 703$ measured images.

\begin{figure}
   \begin{center}
   \begin{tabular}{c}
   \includegraphics[width = 12cm]{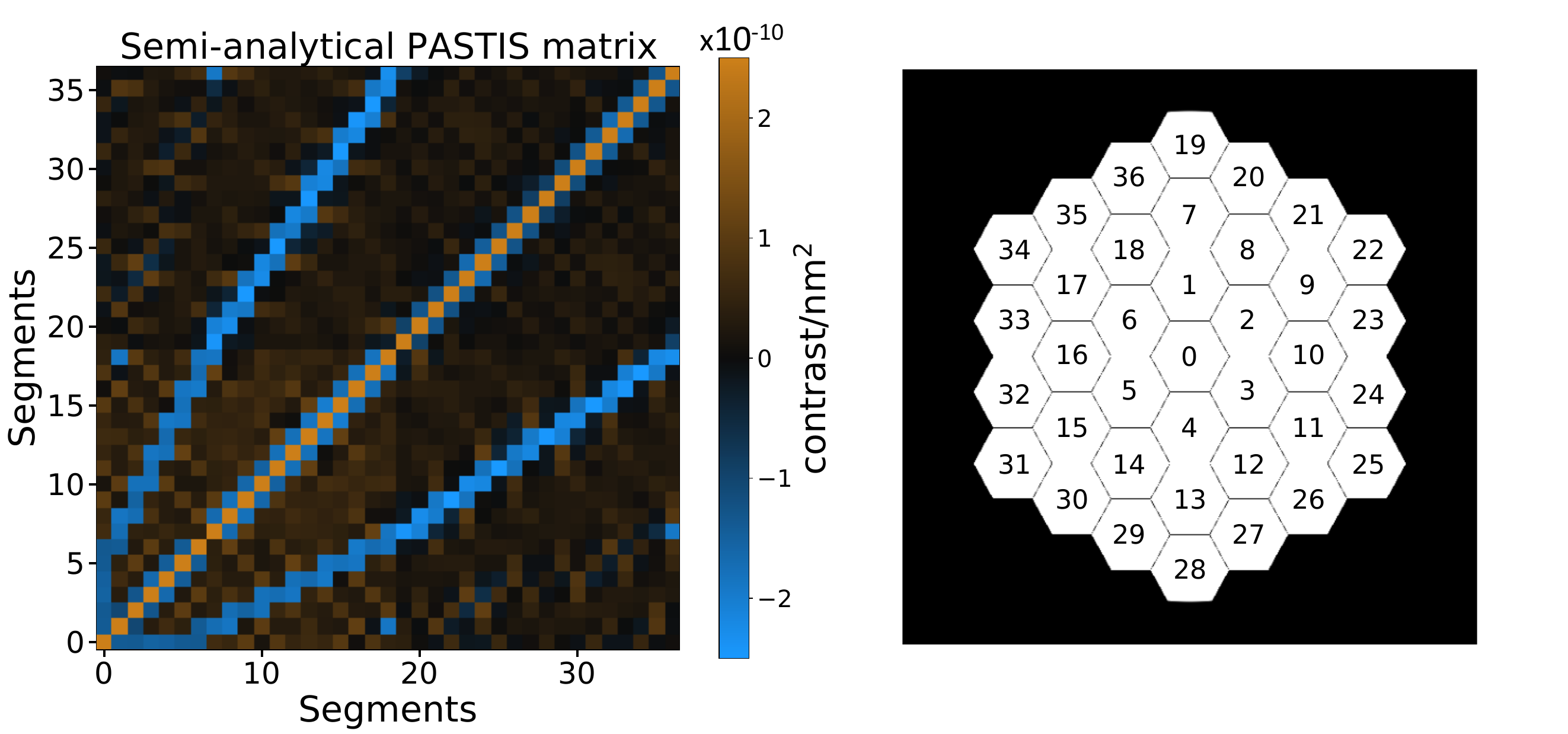}
   \end{tabular}
   \end{center}
   \caption[PASTIS matrix] 
   {\label{fig:pastis-matrix} 
    \textit{Left:} PASTIS matrix for HiCAT as measured in an emulated experiment. Each entry represents the differential contrast contribution of each aberrated segment pair, normalized to the aberration amplitude. The matrix is symmetric, and its diagonal shows the impact on contrast by the individual segments, which is used in the independent segment tolerancing. \textit{Right:} Geometry of the IrisAO segmented DM on HiCAT and the segment numbering used in this paper. The 37 segments are numbered starting at 0 for the center segment, to 36 in the outer ring.}
   \end{figure}

In the presented experiment, we obtain the PASTIS matrix on the HiCAT emulator, and we intend to do exactly the same on the testbed hardware. We constrain ourselves to a local piston mode with an amplitude of $10~nm$ for the calibration aberration of the PASTIS matrix. Other modes are possible, e.g. tip/tilt, or a combination of local segment aberrations, but they are not considered in this paper. Following the semi-analytical approach, we first calculate the contrast matrix by aberrating pairs of segments and recording the resulting DH average contrast. We then use Eqs.~\ref{eq:diagonal-elements} and \ref{eq:off-diagonal-elements} to calculate the PASTIS matrix shown in Fig.~\ref{fig:pastis-matrix}. The PASTIS matrix is fully symmetric, with its diagonal describing the impact on contrast by the individual segments, which is used in the independent segment tolerancing. There are some dark streaks in the matrix with a very low change of contrast for particular segment pairs, which correspond to adjacent segments in the pupil.

\begin{figure}
   \begin{center}
   \begin{tabular}{c}
   \includegraphics[width = 9cm]{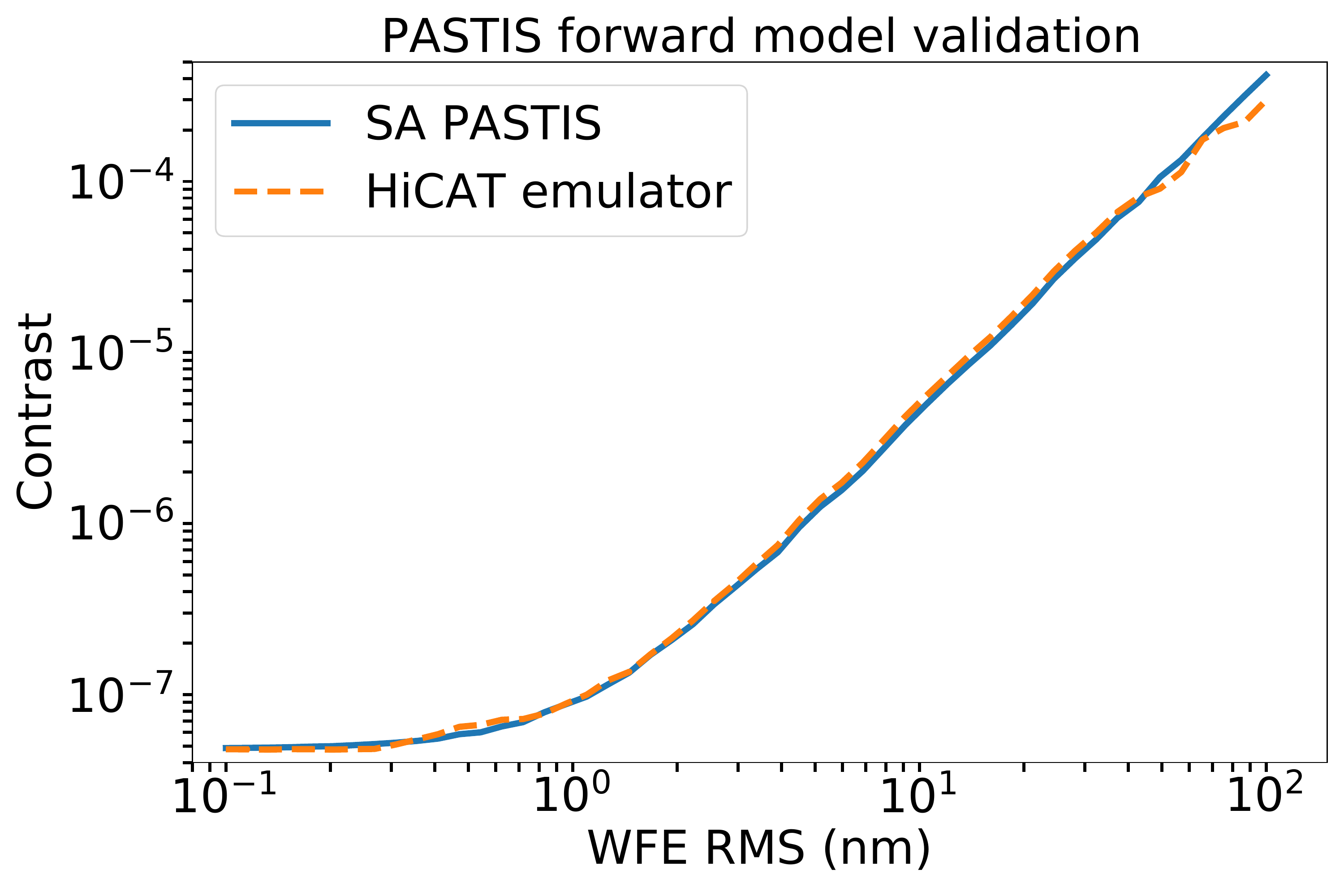}
   \end{tabular}
   \end{center}
   \caption[Hockey stick validation] 
   {\label{fig:hockeyestick} 
    Validation of PASTIS matrix by propagating the same segmented WFE maps both with the semi-analytical PASTIS matrix from Fig.~\ref{fig:pastis-matrix} (solid blue) and with the emulated HiCAT testbed (dashed orange). The curve flattens out to the left, at the coronagraph floor $c_0$, and shows linear behavior at larger WFE, giving it its hockey stick-like shape. The two propagators show very good accordance in the small aberration regime right above the contrast floor, between $\sim 0.6$ and $4~nm$ rms over the entire segmented mirror.}
   \end{figure}

To validate the PASTIS forward model using the PASTIS matrix in Eq.~\ref{eq:pastis-equation}, we generate random segment phase aberrations over the entire segmented DM, $\mathbf{a}$, scale them to a global rms WFE and propagate them with with the PASTIS model. In parallel, we also apply this aberration to the HiCAT segmented DM in simulation and measure the resulting DH contrast. Since one particular rms WFE over the total pupil can be realized with many different individual segment configurations, we average over the contrast values from 10 different realizations at each rms WFE value. The result of this experiment is shown in Fig.~\ref{fig:hockeyestick}. We observe that the two propagators show very good accordance, and more so in the small aberration regime up to $\sim 2~nm$ rms, just above the coronagraph floor, even if the error between them grows only marginally beyond that. We can clearly see this curve flatten out towards the left, where it is limited by the coronagraph floor, and the two solutions start diverging from each other at a global WFE value of around $100~nm$ rms.

\begin{figure}
        \includegraphics[width=1.\linewidth]{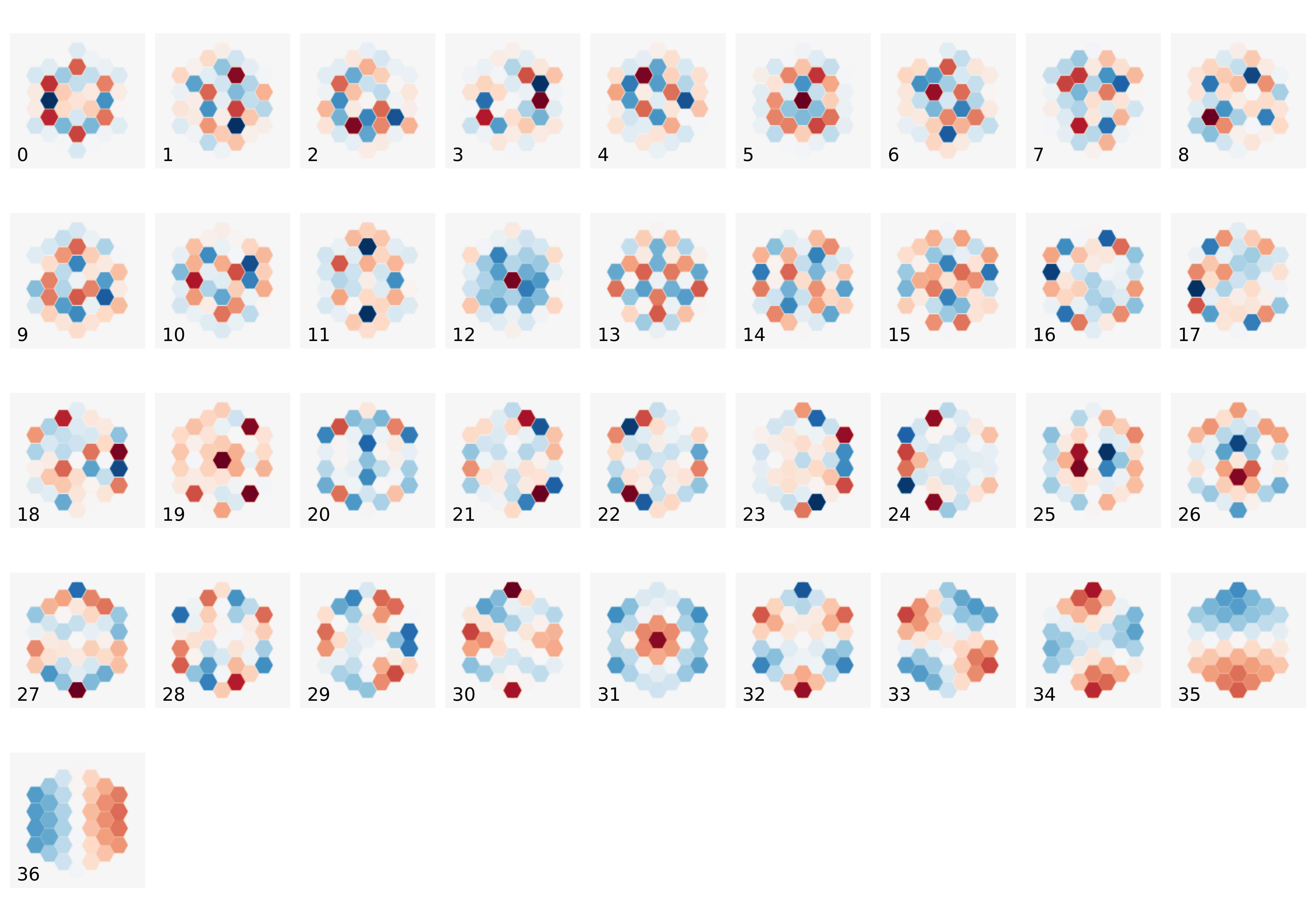}
    \caption{All PASTIS modes for HiCAT with a classical Lyot coronagraph, for local piston aberrations, sorted from highest to lowest eigenvalue. The modes are unitless, showcasing the relative scaling of the segments to each other, and between all modes. They gain physical meaning when multiplied by a mode aberration amplitude $b_p$ in units of wavefront error or phase. Their relative impact on final contrast is given by their eigenvalues.}
    \label{fig:postage-stamp-modes}
    \end{figure}

We proceed with an eigendecomposition of the PASTIS matrix\cite[Sec.~3]{Laginja2020} and calculate its eigenmodes, shown as the PASTIS modes in Fig.~\ref{fig:postage-stamp-modes}. The modes are ordered from highest to lowest eigenvalue, indicating their comparative impact on the DH average contrast in their natural normalization. In Sec.~\ref{subsec:mode-tolerances} we scale these modes to a uniform contrast contribution between them.

\subsection{Validation of mode tolerances}
\label{subsec:mode-tolerances}

\begin{figure}
   \begin{center}
   \begin{tabular}{c}
   \includegraphics[width = 14cm]{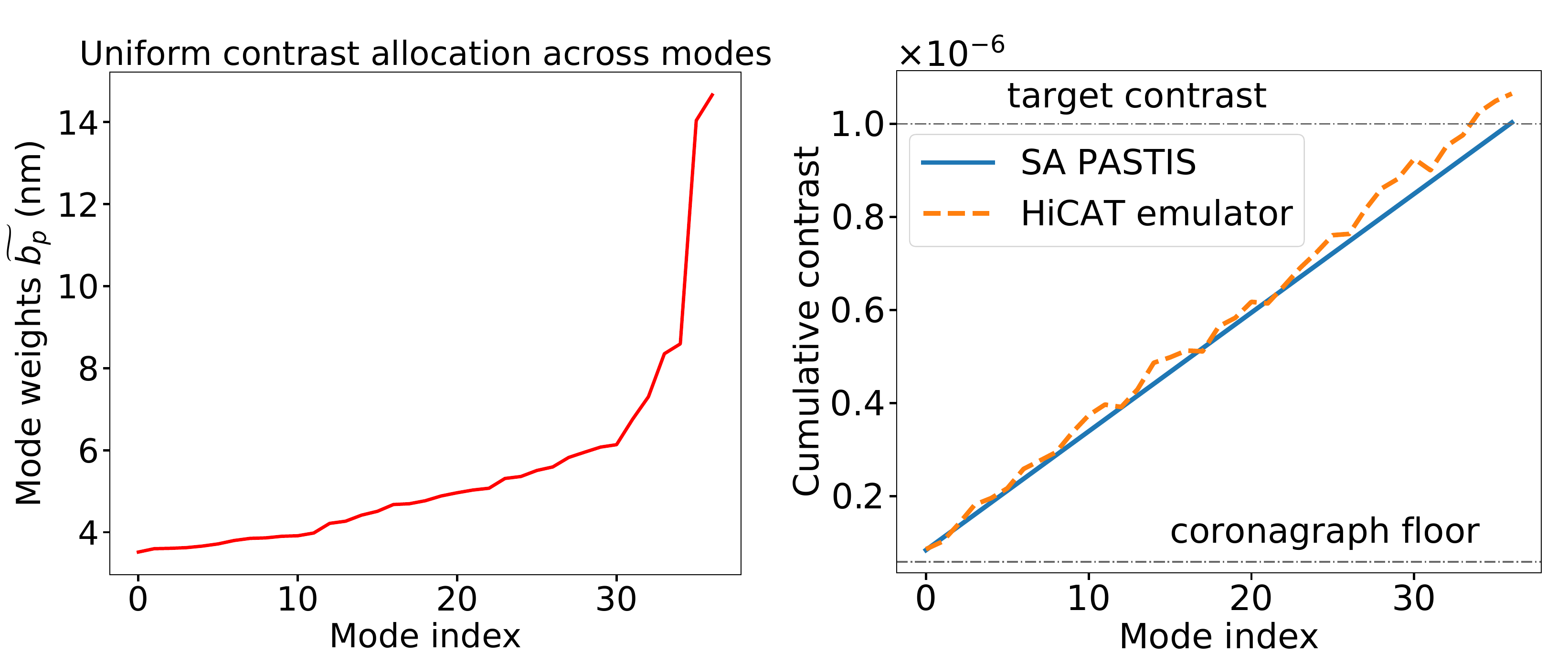}
   \end{tabular}
   \end{center}
   \caption[Cumulative contrast plot] 
   {\label{fig:mode-requirements}
    \textit{Left:} Mode requirements as calculated with Eq.~\ref{eq:calc-sigma-uniform} for a uniform contrast contribution per mode to a target contrast of $10^{-6}$. \textit{Right:} Cumulative contrast plot for the uniform mode requirements shown left, calculated both with the PASTIS propagator (solid blue) and measured with the HiCAT emulator (dashed orange). The experimental result shows a better accordance with the PASTIS model at lower mode index. Note how neither line starts at the coronagraph floor because the mode with index 0 already adds a contrast contribution on top of the baseline contrast.}
   \end{figure}

The PASTIS modes in Fig.~\ref{fig:postage-stamp-modes} form an orthonormal mode basis, making them independent from each other - each of them contributes to the overall contrast without influence from the other modes. This can be used to define error budgets based purely on these optical modes\cite[Sec.~3.2]{Laginja2020}. In the present example, we chose that the PASTIS modes should contribute uniformly to the total contrast, in which case we can calculate the mode tolerances for a particular target contrast with Eq.~\ref{eq:calc-sigma-uniform}. The resulting mode requirements for a target contrast of $c_t = 10^{-6}$ are shown in Fig.~\ref{fig:mode-requirements}, left. To validate the assumption of modes that are independent in contrast, we run an experiment to measure the cumulative contrast of the toleranced PASTIS modes. For this, we multiply the modes by their respective requirement, apply them cumulatively to the IrisAO and measure the resulting DH average contrast at each step (Fig.~\ref{fig:mode-requirements}, right).
The cumulative measurements with the HiCAT emulator follow the general expected linear shape, although some mode contributions seem to overshoot its predicted contrast contribution slightly. These over-contributing modes then seem to be compensated by weighted modes that do not influence the contrast quite as much as intended, displaying a periodic error pattern, and reaching a final contrast 5\% ($5\times10^{-8}$) above the target contrast of $10^{-6}$.

\subsection{Validation of independent segment tolerances}
\label{subsec:segment-tolerances}

\begin{figure}
   \begin{center}
   \begin{tabular}{c}
   \includegraphics[width = 12cm]{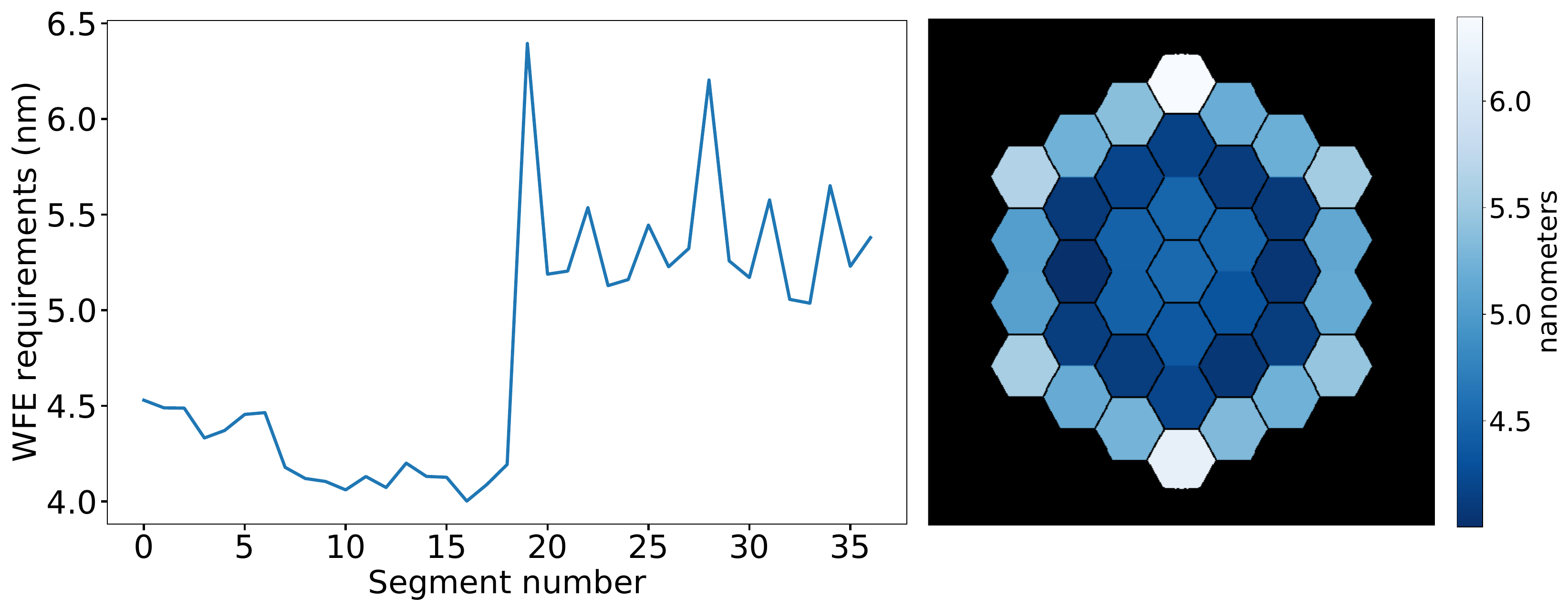}
   \end{tabular}
   \end{center}
   \caption[Segment requirements] 
   {\label{fig:segment-requirements} 
    Independent segment requirements as calculated with Eq.~\ref{eq:single-mus} for a target contrast of $10^{-6}$. These numbers are the standard deviation of the tolerable WFE rms if the target contrast is to be met. The requirement range spans from $4$ to $6.5~nm$, with a clear jump in the outermost ring, which is highly concealed by the Lyot stop (see Fig.~\ref{fig:hicat_pupil_overlaps}).}
   \end{figure}

To fully validate the PASTIS tolerancing model, we calculate segment-level requirements and probe them with HiCAT. In cases where the segments can be assumed to be independent from each other, as is the case for an IrisAO, we can calculate individual segment requirements\cite[Sec.~4.2]{Laginja2020} with Eq.~\ref{eq:single-mus} as a function of the target contrast. While the overall level of WFE requirements will be highly influenced by the Fourier filtering of the FPM, the different segments display a differential tolerance between them, see Fig.~\ref{fig:segment-requirements}, left. These individual segment requirements will be highly influenced by pupil features of the optical system. Looking at their spatial distribution in the HiCAT pupil, we can see in Fig.~\ref{fig:segment-requirements} (right) that the segments of the outer ring have more relaxed requirements than the two inner rings and the center segment. This will be caused, in large part, by the Lyot stop, which is covering a large fraction of the segments in the outer ring because it is undersizing the pupil, which can be seen in Fig.~\ref{fig:hicat_pupil_overlaps}, left.
The segment-level WFE requirements displayed in Fig.~\ref{fig:segment-requirements} present a statistical description of the allowable segment-level WFE if a target contrast of $10^{-6}$ is to be maintained as a statistical mean over many states of the segmented DM. As long as the change of the segment-level WFE on the DM follows a zero-mean normal distribution whose standard deviation per segment is described by the numbers in Fig.~\ref{fig:segment-requirements}, the target contrast will be a recovered as the statistical mean over many such realizations.

\begin{figure}
   \begin{center}
   \begin{tabular}{c}
   \includegraphics[width = 9cm]{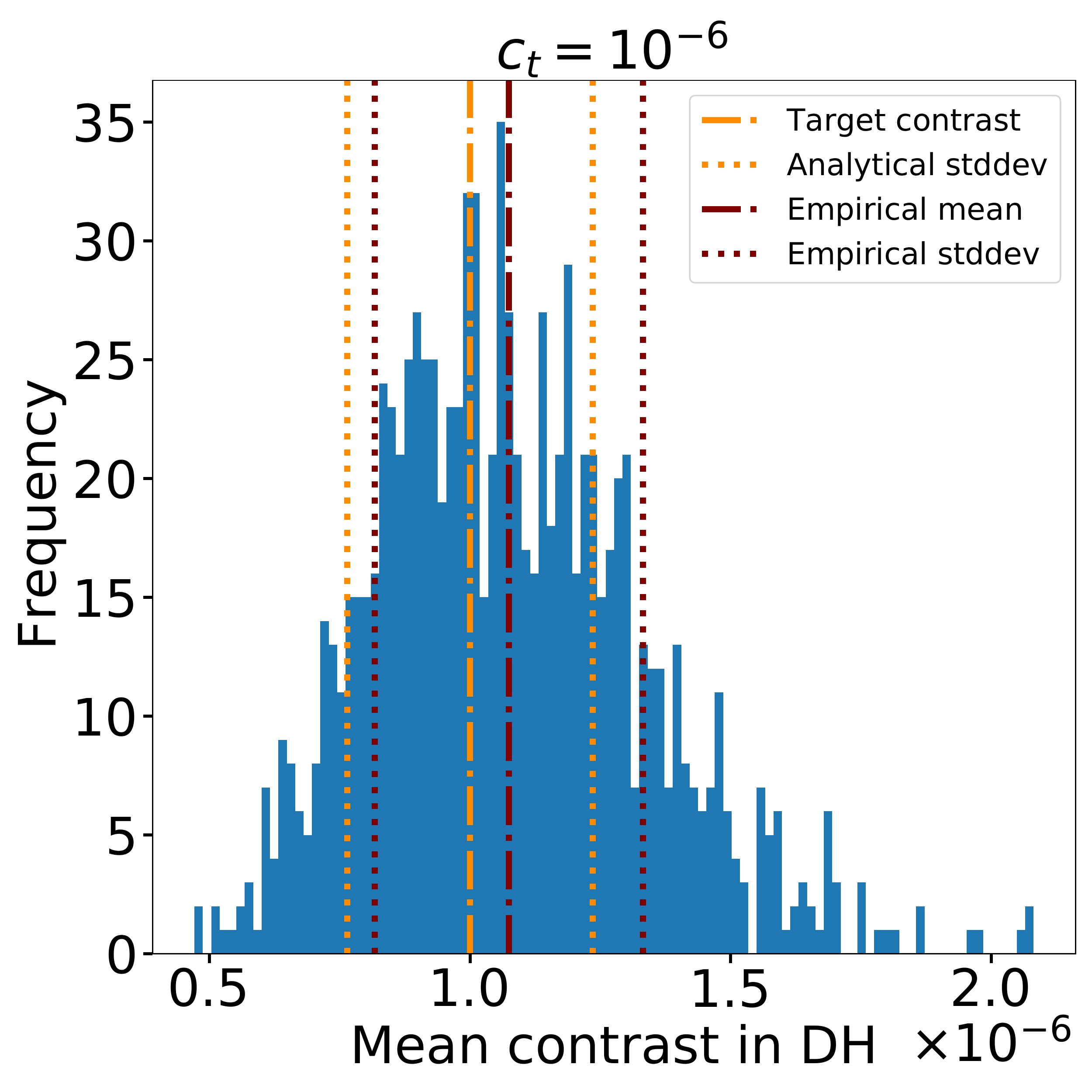}
   \end{tabular}
   \end{center}
   \caption[Monte Carlo segments] 
   {\label{fig:monte-carlo-segments} 
    Monte Carlo validation experiment on the HiCAT emulator to validate the independent segment error budget shown in Fig.~\ref{fig:segment-requirements}, for a target contrast $c_t = 10^{-6}$. Each segmented WFE map draws from 37 zero-mean distributions with an individual standard deviation per segment, $\mu_k$. The expected mean contrast and standard deviation of this distribution, as calculated by Eqs.~\ref{eq:avg-contrast-from-trace} and \ref{eq:var-of-c}, are $\langle c \rangle = 10^{-6}$ (the target contrast) and $2.35 \times 10^{-7}$ (dashed and dotted yellow lines). The measured distribution has a mean of $1.07 \times 10^{-6}$ and a standard deviation of $2.57 \times 10^{-7}$, both of which are larger than the predicted values. This is likely to stem from a combination of a drifting coronagraph floor due to image jitter, and a low number of samples, statistically speaking.}
   \end{figure}

In order to confirm this assumption, we proceed by running a Monte Carlo experiment, producing 1000 different WFE aberration patterns on the segmented DM and recording the propagated average DH contrast. Taking data for 1000 realizations is doable on the hardware in a time frame of about one hour, so we expect to retrieve a histogram with about the same accuracy when performing this experiment on the testbed. The tolerances in Fig.~\ref{fig:segment-requirements} are the prescription as to how to draw the random WFE realizations: each segment-level WFE on segment $k$, in a single random WFE map $\mathbf{a}$, is drawn from its own zero-mean normal distribution with a standard deviation of $\mu_k$. This means that one random HiCAT WFE map is composed of 37 distinct normal distributions with a mean of zero, and a standard deviation of $\mu_k$, which then gets applied to the IrisAO on the HiCAT emulator and propagated through the coronagraph to measure the DH contrast. The distribution of measured average contrast values is shown in Fig.~\ref{fig:monte-carlo-segments}.
The resulting figure corresponds to a Gaussian distribution with a mean of $1.07 \times 10^{-6}$ and a standard deviation of $2.57 \times 10^{-7}$, marked in the plot with dark red lines. To interpret the results in Fig.~\ref{fig:monte-carlo-segments}, we remember that PASTIS provides analytical expressions to derive the expected mean contrast (Eq.~\ref{eq:avg-contrast-from-trace}) and variance (Eq.~\ref{eq:var-of-c}) from a distribution calculated with a set of segment requirements. Apart from the PASTIS matrix $M$, what is needed to calculate these quantities is the segment covariance matrix $C_a$, which in the case of independent segments as presented here is a simple diagonal matrix made of the segment requirement variances, $\mu^2_k$, which we take from Fig.~\ref{fig:segment-requirements}. Then, Eq.~\ref{eq:var-of-c} yields an analytical standard deviation of $2.35 \times 10^{-7}$ and the mean is, as expected, the target contrast value $10^{-6}$. These are marked with yellow lines in Fig.~\ref{fig:monte-carlo-segments}. We observe that both the measured mean as well as the variance are higher than the analytically calculated values from the PASTIS matrix. This discrepancy could be attributed to a drifting coronagraph floor, as the baseline contrast will not be exactly the same like the one that was used in the calculation of the segment tolerances; it could also be a result of small-sample statistics.

Overall, our experiments on the HiCAT emulator present a successful implementation of empirical validations of the PASTIS model for a specific high contrast instrument, the HiCAT testbed. We measured a PASTIS matrix and validated it by comparing its propagation results with measurements from the synthetic testbed. We decompose the matrix into independent optical modes that we toleranced uniformly and cumulatively to a target contrast of $10^{-6}$. Finally, we calculate segment-level WFE requirements under the assumption of independent segments and validate them with a Monte Carlo experiment, measuring the contrast from randomly drawn segmented WFE maps as prescribed by the derived requirements.


\section{CONCLUSIONS AND FUTURE WORK}
\label{sec:CONCLUSIONS}

The PASTIS tolerancing model is well established\cite{Laginja2020, Leboulleux2018jatis}: it includes a semi-analytical coronagraphic propagation model for segmented phase aberrations, and a statistical framework for the calculation of segment-based WFE requirements. This model has been validated against an end-to-end simulator for the LUVOIR-A case, confirming the properties of the optical eigenmodes and derived segment-level tolerances.

The most recent results on the HiCAT testbed show a contrast performance of $\sim5\times10^{-8}$ in a 6-11 $\lambda/D_{LS}$ circular DH with pair-wise and stroke minimization, in an unsegmented CLC configuration. The recent addition of an IrisAO segmented DM on hardware makes it the ideal environment for the validation of PASTIS on an actual high contrast instrument, at moderate contrasts between $10^{-6}$ and $10^{-8}$ in ambient conditions. In order to prepare for these empirical validations, we implemented a set of experiments on the HiCAT testbed emulator that include the newly installed segmented DM, with realistic testbed residual WFE and noise. These emulated experiments are embedded in the fully functional HiCAT control infrastructure, which will let us run them on hardware as soon as the testbed is fully calibrated.

The results we present show three experiments that we want to perform on the hardware testbed:
    \begin{enumerate}
    \item Measure an empirical PASTIS matrix and validate the PASTIS forward model against hardware measurements.
    \item Validate the individual contrast contributions from the scaled PASTIS modes.
    \item Validate the independent segment-level requirements as calculated with PASTIS through a Monte Carlo experiment.
    \end{enumerate}
We have shown that we can perform these empirical tests on HiCAT in the $10^{-6}$ to $10^{-7}$ contrast range, with the results showing a successful validation of the PASTIS model tolerancing predictions. The measured PASTIS matrix is used for a set of propagations of random segment aberrations and successfully validated against equal propagations on the testbed emulator. The matrix was then decomposed into eigenmodes which we scale uniformly to yield a cumulative target contrast of $10^{-6}$. We proceeded by measuring the cumulative contrast impact from these toleranced optical modes. Using the same target contrast, we calculate the independent segment requirements which showed to be WFE standard deviations in a range between $4$ and $6.5~nm$ rms. We use these to draw random segmented WFE maps from zero-mean normal distributions, measuring the resulting average DH contrast for 1000 random realizations. The empirical mean of the resulting distribution recovers the target contrast to a small error.

The experiments we prepared on the emulator show that we might anticipate certain challenges when moving to the real hardware. The limitations imposed by the IrisAO will define how well we can control the aberration modes we introduce. While this segmented DM is known to have close to perfect linear behavior, the modes applied to it will stem from open loop calibrations, which might include some errors. Furthermore, the least significant bit introduced by the IrisAO controller will prevent us from aberrating a segment with an amplitude smaller than $\sim1~nm$ when taking noise into account. This will lead to drawing from imperfect normal distributions when creating random WFE maps with the prescription of the segment requirements, truncating any small aberrations to zero, which might skew the results of the Monte Carlo analysis. However, we expect these errors to be small if the target contrast is chosen high enough compared to the coronagraph floor. Another challenge will be the stability of the DH solution that we adopt into our coronagraph, setting the contrast floor we perform our experiments at. Since the independent segment tolerances $\mu_k$ are a function of the contrast floor (Eq.~\ref{eq:single-mus}), a drifting baseline contrast during an experiment will likely introduce a constant shift in the final results. Recent work on HiCAT reduced this drift significantly, enabling us to recover the DH contrast of $5-6\times10^{-8}$, reached by a WFS\&C loop, many times over a period of two weeks simply by invoking the same DM commands, to an error of $0.5-1\times10^{-7}$. We intend to mitigate the residual effects of this by running our experiments on timescales that are faster than the drifts, and by regularly measuring the effective contrast floor where possible.

The apparent next step is to perform these experiments on the HiCAT testbed. In the future, we intend to explore the usage of the PASTIS modes as part of the low-order wavefront sensor, determining the sensitivity of the sensor to these modes, and to investigate options for modal control.

\acknowledgments 
This work was co-authored by employees of BALL AEROSPACE as part of the the Ultra-Stable Telescope Research and Analysis (ULTRA) Program under Contract No. 80MSFC20C0018 with the National Aeronautics and Space Administration (PI: L. Coyle), and by STScI employees under corresponding subcontracts No.18KMB00077 and No.19KMB00102 with Ball Aerospace (PI: R. Soummer, Sci-PI: L. Pueyo). This work was also co-authored by employees of the French National Aerospace Research Center ONERA (Office National d'\'{E}tudes et de Recherches A\'{e}rospatiales), and benefited from the support of the WOLF project ANR-18-CE31-0018 of the French National Research Agency (ANR), as well as the internal research project VASCO. This work was also supported in part by the National Aeronautics and Space Administration under Grant 80NSSC19K0120 issued through the Strategic Astrophysics Technology/Technology Demonstration for Exoplanet Missions Program (SAT-TDEM; PI: R. Soummer), and by the Segmented-aperture Coronagraph Design and Analysis funded by ExEP, under JPL subcontract No.1539872, and by the STScI Director's Discretionary Research Funds.
The United States Government retains and the publisher, by accepting the article for publication, acknowledges that the United States Government retains a non-exclusive, paid-up, irrevocable, worldwide license to reproduce, prepare derivative works, distribute copies to the public, and perform publicly and display publicly, or allow others to do so, for United States Government purposes. All other rights are reserved by the copyright owner. 

This research was developed in Python\footnote{\url{https://www.python.org}}, an open source programming language. HiCAT makes use of the Numpy\cite{Oliphant2006numpy, vanderWalt2011numpy}, Matplotlib\cite{Hunter2007matplotlib, matplotlib_v3.3.3}, Astropy\cite{AstropyCollaboration2013, AstropyCollaboration2018, Astropy2018zenodo}, SciPy\cite{scipy}, scikit-image\cite{scikit-image}, pandas\cite{Mckinney2010pandas, pandas_v1.1.4}, imageio\cite{imageio}, photutils\cite{photutils_v.1.0.1} and HCIPy\cite{Por2018hcipy} packages. The software for the tolerancing analysis used PASTIS\cite{Laginja2020pastis}, a modular, open-source Python package for segment-level error budgeting of segmented telescopes.

\bibliography{2020jatis}
\bibliographystyle{spiebib}

\end{document}